\documentstyle[a4,11pt,epsf]{article}
\epsfverbosetrue
 0

\newcommand{\beq}{\begin{equation}}
\newcommand{\eeq}{\end{equation}}

\begin{document}

\begin{titlepage}
 
\vspace{5mm}
 
\begin{center}
{
 \huge 
       Exact Finite-Size Scaling and Corrections
  \\[3mm]
 to Scaling in the Ising
 Model with
  \\[7mm]
 Brascamp-Kunz Boundary Conditions
}
\\[15mm]
{\bf 
W. Janke$^{\rm{a}}$ 
and 
{R. Kenna$^{\rm{b}}$},
\\
$^{\rm{a}}$ 
Institut f\"ur Theoretische Physik, Universit\"at Leipzig,
\\Augustusplatz 10/11, 04109 Leipzig, Germany\\
$^{\rm{b}}$ 
School of Mathematics, Trinity College Dublin, Ireland
} 
\\[3mm]~\\ 
March 2001
\end{center}
\begin{abstract}
The Ising model in two dimensions with the special boundary
conditions of Brascamp and Kunz is analysed. 
Leading and sub-dominant  scaling behaviour of the Fisher 
zeroes are determined exactly.
The finite-size scaling, with corrections,
 of the specific heat is determined 
both at the critical and pseudocritical points.
The shift exponents associated with scaling of the pseudocritical 
points are not the same as the inverse correlation length
critical exponent. 
All corrections to scaling  are analytic. 
\end{abstract}
\end{titlepage}

\newpage

\section{Introduction}
\label{introduction}
\setcounter{equation}{0}

Second order critical phenomena are signaled by
the divergence of an appropriate second derivative of 
the free energy along with the correlation length.
For temperature driven phase transitions, this divergence
is in the specific heat.
Of central interest in the study of such phenomena is the 
determination of the critical exponents which characterize
these divergences. Here, the concept of
universality plays a fundamental role. 
The universality hypothesis asserts that critical behaviour
is determined solely 
by the number of space (or space time) dimensions
and by the symmetry properties of the order parameters
of the model.
The universality class is thus labeled by critical exponents
describing the singular behaviour of thermodynamic functions
in the infinite volume limit.
In finite systems the counterparts of these singularities 
are smooth peaks the shapes of which depend on the critical 
exponents. 
Finite-size scaling (FSS) is a well established technique
for  the numerical or analytical
extraction of these
exponents from finite volume analyses \cite{Fi71,Barber,Privman,JaKe01}.

In particular, 
let $C_L (\beta)$ be the specific heat at inverse temperature $\beta$
for a system of linear extent $L$.
FSS of the specific heat is
characterized by {\em{(i)}} the location of its peak,
$\beta_L$, {\em{(ii)}} its height $C_L(\beta_L)$ and 
{\em{(iii)}} its value at the infinite volume critical point 
$C_L(\beta_c)$. The position of the specific heat peak, $\beta_L$, is a 
pseudocritical point which approaches $\beta_c$ as $L 
\rightarrow \infty$ in a manner dictated by the shift exponent $\lambda$,
\begin{equation}
 |\beta_L - \beta_c| \sim L^{-\lambda}
\quad .
\end{equation}
Finite-size scaling theory also gives that if the specific heat 
divergence is of a power-law type in the thermodynamic limit,
$C_L \sim |\beta - \beta _c|^{-\alpha}$, then its peak
behaves with $L$ as
\begin{equation}
 C_L(\beta_L) \sim L^{\alpha /\nu}
\quad ,
\label{FSSCv}
\end{equation}
in which $\nu$ is the correlation length critical exponent.
When $\alpha=0$, which is the case in the Ising model in two
dimensions,  this behaviour  is modified to
\begin{equation}
 C_L(\beta_L) \sim \ln{L}
\quad .
\end{equation}
In most models the shift exponent $\lambda$ coincides with $1/\nu$, 
but this is not a direct conclusion of finite-size scaling and 
is not always true.
If the shift exponent obeys $\lambda \ge 1/\nu$
then (\ref{FSSCv}) also holds if the pseudocritical temperature
$\beta_L$ is replaced by the critical one $\beta_c$.
If, on the other hand, $\lambda < 1/\nu$, one has
$C_L(\beta_c) \sim L^{\lambda \alpha}$ in the power-law case \cite{Barber}.

The leading FSS behaviour of a wide range of models is by now
well understood. 
Recently, attention has focused on the determination of
corrections to scaling   
\cite{GoMa93,DGS,HoLa96,ADH,boris,HuCh99,WuLu01,KaOk01}.
These corrections may arise from irrelevant scaling fields or be
analytic in $L^{-1}$.
Typically, numerically based 
or experimental studies 
involve systems of limited size where these corrections to
leading FSS behaviour cannot be dismissed.
A better knowledge of universal sub-dominant behaviour would 
therefore be of great benefit in FSS extrapolation procedures
\cite{Kaufman,FF,Beale,CaPe98,SaSa00,Qu00,IzHu00,Sa00,PeVi00,IvIz01}.

The Ising model in two dimensions is the simplest statistical physics
model displaying critical behaviour. Although solved in the absence
of an external field \cite{solvedIsing}, 
it remains a very useful testing ground in 
which new techniques can be explored in the hope of eventual 
application to other, less understood, models in both statistical
physics and in lattice field theory. 

Exploiting the exactly known partition function of the
two dimensional Ising model on finite lattices with toroidal
boundary conditions \cite{Kaufman},
Ferdinand and Fisher \cite{FF} analytically determined the 
specific heat FSS to order $L^{-1}$.
At the infinite volume critical point
this was recently extended  to order $L^{-3}$ by 
Izmailian and Hu \cite{IzHu00}
and independently by Salas \cite{Sa00}.
It was found that only integer powers 
of $L^{-1}$ occur, with no logarithmic modifications (except of 
course for the leading logarithmic term), i.e.,
\begin{equation}
 C_L(\beta_c) =
 C_{00}\ln{L} + C_0
 + \sum_{k=1}^\infty{\frac{C_k}{L^k}}
\quad .
\label{Salas2.13}
\end{equation}
For the toroidal lattice, the coefficients $C_{00}$, $C_0$,
$C_1$,
$C_2$ and $C_3$ have been  determined explicitly 
\cite{FF,IzHu00,Sa00,solvedIsing}.
Ferdinand and Fisher also determined the behaviour
of the specific heat
pseudocritical point, finding $\lambda = 1 = 1/\nu$
(except for special values of the ratio of the lengths of the lattice
edges, in which case  pseudocritical scaling was found to be of the form
$L^{-2}\ln{L}$).

The specific heat of the Ising model has also recently been studied 
numerically on two
dimensional lattices with other boundary conditions in 
Refs.~\cite{DGS,HoLa96}. 
For lattices with spherical 
topology, the correlation length and shift exponents
were found to be 
$\nu = 1.00 \pm 0.06$ and 
$\lambda=1.745\pm0.015$, significantly away from
$1/\nu$ \cite{DGS}.
This is compatible with
an earlier study  reporting $\lambda \approx
1.8$ \cite{GoMa93}. 
Therefore the FSS of the specific heat pseudocritical point 
does not match the correlation length scaling behaviour.
This is in contrast to the situation with toroidal boundary conditions,
where $\lambda=1/\nu = 1$ \cite{FF}.
On the other hand, it was established in
\cite{DGS} that the
critical properties on such a lattice are the same as for the
torus. 

The finite size behaviour of the specific heat
is related to that of the complex temperature zeroes of the
partition function, the so-called
Fisher zeroes \cite{Fi64}.
Indeed, the FSS of the latter provides further information on
the critical exponents of the model.
The leading FSS behaviour of the imaginary part of a Fisher zero
is \cite{IPZ}
\begin{equation}
 {\rm{Im}}z_j(L) \sim L^{-1/\nu}
\quad ,
\label{IPZI}
\end{equation}
while the real part of the lowest zero 
may be viewed as another pseudocritical point, scaling as
\begin{equation}
  |\beta_c - {\rm{Re}}z_1(L)|\sim L^{-\lambda_{\rm{zero}}}
\quad .
\label{IPZR}
\end{equation}

Hoelbling and Lang used
a variety of cumulants as well as Fisher zeroes
to study universality of the Ising model on sphere-like lattices
\cite{HoLa96}.
They reported
that the imaginary part of the first zero is an optimal quantity 
to determine the exponent $\nu$ which is in perfect agreement 
with unity for all lattices studied. 
Regarding the FSS of the pseudocritical point, the numerical
results on a torus are in agreement with 
$\lambda=1/\nu = 1$ \cite{FF}.
The pseudocritical point (from specific heat, the  real part of the lowest
zero as well as from two other types of cumulant) is not,
however, of the Ferdinand-Fisher type. I.e., the amplitude of
any ${\cal{O}}(1/L)$ contribution is compatible with zero.
While a shift exponent $\lambda = 1.76(7)$ is consistent with 
their numerical results, Hoelbling and Lang point out that
this is not stringent and pseudocritical FSS of the form
$L^{-2}\log{L}$  or even $L^{-2}$ are also compatible with their data. Thus,
while a possible leading $L^{-1}$ term almost or completely
vanishes and subleading terms are dominant, the precise nature
of these corrections could not  be unambiguously decided.
In any case, FSS of the position of the specific heat peak does
not accord with the correlation length exponent for spherical lattices
and the thermodynamic limit 
is achieved faster there than on a toroidal lattice.

In another recent study \cite{ADH} involving Fisher zeroes, 
Beale's \cite{Beale} exact distribution function for the energy 
of the two dimensional Ising model
was exploited
to obtain the exact zeroes
for square periodic lattices up to size $L=64$.
The FSS analysis in \cite{ADH} yielded a value for the
correlation length critical exponent, $\nu$, which appeared
to approach the exact value (unity) as the thermodynamic
limit is approached. Small lattices appeared to yield a 
correction-to-scaling
exponent in broad agreement with early estimates ($\omega \approx 1.8$
\cite{Bi81}).
However, closer to the thermodynamic limit, these corrections 
appeared to be analytic with $\omega=1$.
On the other hand, in \cite{SaSa00}, 
the scaling behaviour of the susceptibility and related quantities was
considered and evidence for $\omega=1.75$ or, possibly, 
 $2$ was presented.

In the light of these recent analyses, we wish to present analytic
results which may clarify the situation. To this end, we have selected
the Ising model with special boundary conditions due to
Brascamp and Kunz \cite{BrKu74}.
These boundary
conditions permit an analytical approach to the determination
of a number of thermodynamic quantities.
Recently, Lu and Wu exploited this fact to determine the density
of Fisher zeroes in the thermodynamic limit \cite{WuLu00}.
In this paper, we take a complimentary approach, exploiting
FSS behaviour (i) to determine critical exponents, (ii) to determine
corrections to leading scaling and (iii) to gain experience
in the hope of eventual application to other, less transparent scenarios.
The rest of this paper is organised as follows.
In Section~2 the Brascamp-Kunz boundary conditions
are introduced and the exact FSS of the Fisher zeroes calculated.
The specific heat and its pseudocritical point are analysed
in  Section~3. Our conclusions are 
contained in Section~4 and the Appendix contains some calculations 
of relevance to the specific heat analysis of Section~3.

\section{The Fisher Zeroes for Brascamp-Kunz Boundary Conditions}
\label{FZ}
\setcounter{equation}{0}

Brascamp and Kunz introduced special boundary conditions, for which the
Fisher zeroes are known for any finite size lattice \cite{BrKu74}. 
They considered a 
regular lattice with  $M$ sites in the $x$ direction and $2N$ sites in 
the $y$ direction. The special boundary conditions are periodic in
the $x$ direction and Ising spins
fixed to $\dots +++ \dots$ and 
$\dots +-+-+- \dots$ along the edges in the $y$ direction.
For such a lattice, the Ising partition function can be rewritten as 
\begin{equation}
 Z_{M,2N} = 2^{2MN}
 \prod_{i=1}^N{
 \prod_{j=1}^M{
 \left[
         1 + z^2 - z (\cos{\theta_i} + \cos{\phi_j})
 \right]
}}
\quad,
\label{ZM2N}
\end{equation}
where $z=\sinh{2 \beta}$, $\theta_i = (2i-1)\pi/2N$ 
and $\phi_j = j \pi/(M+1)$ and where $\beta = 1/k_B T$ is
the inverse temperature. 
The multiplicative form of (\ref{ZM2N}) is of central importance to 
this paper.

Brascamp and Kunz showed that the zeroes of the partition function 
(\ref{ZM2N}) are
located on the unit circle in the complex $z$ plane (so that the critical
point is $z=z_c = 1$). These are 
\begin{equation}
  z_{ij} = \exp{(i \alpha_{ij})}
\quad,
\label{zi}
\end{equation}
where
\begin{equation}
 \alpha_{ij}
 =
 \cos^{-1}{\left( \frac{\cos{\theta_i}+\cos{\phi_j}}{2} \right)}
\quad.
\label{zijM2N}
\end{equation}
Setting $2N=\sigma M$ and using a computer algebra system such as Maple, 
one may expand (\ref{zijM2N}) in $M$ to determine FSS of any
zero to any desired order. Indeed, the first few terms in the expansion
for the first zero which is the one of primary interest, are
\begin{eqnarray}
\lefteqn{
\alpha_{11} =
 M^{-1} \frac{\pi \sqrt{1+\sigma^2}}{\sqrt{2} \sigma } 
 -
 M^{-2} \frac{\pi\sigma }{ \sqrt{2}\sqrt{1+\sigma^2}}
}
&   & 
\nonumber
\\
& & 
 +
 M^{-3} \frac{\pi\sqrt{2}}{(1+\sigma^2)^{5/2}}
 \left\{
   \frac{\sigma}{4}(1+\sigma^2)(3+2\sigma^2)
   -
   \frac{\pi^2}{96\sigma^3}(1-\sigma^4)^2
 \right\}
\nonumber
\\
& & 
 -
M^{-4}
\frac{ \pi \sqrt{2}}{\sigma(1+\sigma^2)^{5/2}}
\left\{
       \frac{\sigma^2}{4}
       (4+5\sigma^2+2\sigma^4)
       +
       \frac{\pi^2}{96}
       (5+3\sigma^2)
       (1-\sigma^2)
       (1+\sigma^2)
\right\}
\nonumber
\\
& &  + {\cal{O}}\left(M^{-5}\right)
\quad.
\label{aa}
\end{eqnarray}
Separating out the real and imaginary parts of the FSS of the first zero
yields
\begin{eqnarray}
\lefteqn{
 {\rm{Re}}z_{11} = \cos{\alpha_{11}} =
}
&   & 
\nonumber
\\
& & 
 1
 -
 M^{-2} \frac{\pi^2}{4} \left( 1+ \frac{1}{\sigma^2}  \right)
 +
 M^{-3} \frac{\pi^2}{2} 
 +
 M^{-4} \frac{\pi^2}{4} 
  \left( -3 + \frac{\pi^2}{12}\left( 
                                     1+ 
                                   \frac{1}{\sigma^4} 
                              \right)
 \right)
 +
 {\cal{O}}\left(M^{-5}\right)
\quad ,
\label{bb}
\end{eqnarray}
and
\begin{eqnarray}
\lefteqn{
{\rm{Im}}z_{11} =  \sin{\alpha_{11}} =
\frac{ \pi \sqrt{2}}{\sigma (1+\sigma^2)^{5/2}}
\left\{
 M^{-1}\frac{(1+\sigma^2)^3}{2}
 -
 M^{-2}\frac{\sigma^2(1+\sigma^2)^2}{2}
\right.
}
&   & 
\nonumber
\\
& & 
\left.
 +
 M^{-3}
 \frac{1}{96\sigma^2}
 \left[
        -\pi^2 (1+\sigma^2)^2 
       \left(
             5+6\sigma^2+5\sigma^4
       \right)
       +
       24 \sigma^4 (1+\sigma^2) (3+2\sigma^2) 
 \right]
\right.
\nonumber
\\
& & 
\left.
 +
M^{-4}
 \left[
       -\frac{\sigma^2}{4} \left(4+5\sigma^2
       +2\sigma^4 \right)
        + 
       \frac{\pi^2}{96}
        (1+\sigma^2) 
        (1+3\sigma^2) 
        (7+5\sigma^2) 
 \right]
\right\}
       +
 {\cal{O}}\left(M^{-5}\right)
\quad ,
\label{cc}
\end{eqnarray}
respectively.
 From the leading term in (\ref{cc}) and from (\ref{IPZI}) one has,
indeed, that the correlation length critical exponent is $1$. 
Note, however, from (\ref{bb}) that the leading FSS behaviour
of the pseudocritical point in the form of the real part of the lowest 
zero is
\begin{equation}
 z_c - {\rm{Re}}z_{11} = 1 - {\rm{Re}}z_{11}
 \sim M^{-2}
\quad ,
\label{psze}
\end{equation}
giving shift exponent $\lambda_{{\rm{zero}}} = 2$. This value is consistent 
with the numerical results of \cite{HoLa96} for spherical lattices.
One further notes that all corrections are powers of $M^{-1}$ and
in this sense entirely analytic.

Setting $\sigma=1$ gives the FSS behaviour of the first  zero
for a square $M\times M$ lattice to be
\begin{equation}
 \alpha_{11}
 =
 \pi M^{-1}
 - \frac{\pi}{2}M^{-2}
 + \frac{5 \pi}{8}M^{-3}
 -\frac{11 \pi}{16}M^{-4}
 + {\cal{O}}\left(M^{-5}\right)
\quad,
\end{equation}
which, for the real and imaginary parts separately is
\begin{eqnarray}
 {\rm{Re}}z_{11} & = & \cos{\alpha_{11}}
 =
 1
 -\frac{\pi^2}{2}M^{-2}
 +\frac{\pi^2}{2}M^{-3}
 -\frac{\pi^2}{4}\left( 3 - \frac{\pi^2}{6} \right) M^{-4}
 +{\cal{O}}\left(M^{-5}\right)
\quad ,
\label{Rez}
\\
 {\rm{Im}}z_{11} & = & \sin{\alpha_{11}}
 = 
 \pi M^{-1}
 -\frac{\pi}{2}M^{-2}
 -\frac{\pi}{2}\left( \frac{\pi^2}{3} - \frac{5}{4} \right) M^{-3}
\nonumber \\
& & 
\quad \quad \quad\quad \quad \quad \quad \quad \quad \quad \quad \quad 
 +\frac{\pi}{4}\left( \pi^2 - \frac{11}{4}\right) M^{-4}
 +{\cal{O}}\left(M^{-5}\right)
\quad .
\label{Imz}
\end{eqnarray}

In summary, we have observed that for the first zero, the shift
exponent is not $1/\nu$ and that all corrections
are analytic. 
Expansion of higher zeroes yields the same result.

\section{The Specific Heat}
\setcounter{equation}{0}
In terms of the variable $z=\sinh{2\beta}$, the specific heat is
\begin{equation}
 C \equiv \frac{k_B \beta^2}{V} \frac{\partial^2 \ln{Z}}{\partial \beta^2}
= \frac{4  k_B \beta^2}{V} \left[
 (1+z^2)\frac{\partial^2 \ln{Z}}{\partial z^2}
 +
 z \frac{\partial \ln{Z}}{\partial z}
\right]
\quad ,
\label{cful}
\end{equation}
where $V$ is the volume of the system.
The singular behaviour comes from the first term only, the second
being entirely regular. Thus we may split (\ref{cful}) into
singular and regular parts and retain only the former.
 From (\ref{ZM2N}),  this singular part of the specific heat 
for an $M \times 2N$ lattice
is (up to the factor $(1+z^2) 4  k_B \beta^2$)
\begin{equation}
 C_{M,2N}^{\rm sing.}(z) = 
 \frac{1}{2MN}
 \sum_{i=1}^N{
 \sum_{j=1}^M{
 \left\{
           \frac{2}{1+z^2-z(\cos{\theta_i}+\cos{\phi_j})}
           -
           \frac{[2z-(\cos{\theta_i}+\cos{\phi_j})]^2
                }{
           [1+z^2-z(\cos{\theta_i}+\cos{\phi_j})]^2}
 \right\}
}
}
\quad .
\label{CM2N}
\end{equation}
An alternative approach is to write the partition function as
\begin{equation}
 Z_{M,2N} = A(z) \prod_{i=1}^N \prod_{j=1}^M(z-z_{ij})(z-{\bar{z}}_{ij}) 
\quad ,
\end{equation}
where the product is taken over all zeroes.
The non-vanishing function $A(z)$ contributes only to the regular
part of the partition function, and from 
(\ref{zi}) and (\ref{zijM2N}), the product
term leads precisely to the singular 
expression (\ref{CM2N}).

It is straightforward to perform the $i$- summation first in (\ref{CM2N}).
Indeed, (\ref{Ia}) and  (\ref{Ib}) yield the exact result, which
is conveniently expressed as
\begin{equation}
 C_{M,2N}^{\rm sing.}(z)
 =
 \frac{1}{z^3}
 \frac{1}{M}
 \sum_{j=1}^M{g_j(z)}
 +
 \left(
       1 - \frac{1}{z^2}
 \right)
 \frac{1}{2M}
 \sum_{j=1}^M{g_j^{(1)}(z)}
 - \frac{1}{2z^2}
\quad ,
\label{ce}
\end{equation}
where
\begin{equation}
 g_j(z)
 =
 \frac{
       \tanh{
             \left(
                    N
                    \cosh^{-1}{
                               (1/z+z-\cos{\phi_j})
                              }
             \right)
            }
     }{
       \sqrt{
             (1/z+z-\cos{\phi_j})^2-1
            }
     }
\quad ,
\end{equation}         
and where $g_j^{(k)}(z) = d^k g_j(z) / dz^k$.

It is convenient at this stage to introduce the sums
\begin{equation}
 S_k =
 \frac{1}{M}
 \sum_{j=1}^M g_j^{(k)}(1)
\quad,
\end{equation}
and to make the observation that
\begin{eqnarray}
 g_j^{(1)}  (1) & = & 0
\quad ,
\label{simpl1}
\\
 g_j^{(3)}  (1) & = & - 3 g_j^{(2)}(1)
\quad.
\label{simpl2}
\end{eqnarray}

\paragraph{Specific Heat at the Critical Point:}

At the infinite volume critical temperature, $z=z_c=1$,
the specific heat is, from (\ref{ce}),
\begin{equation}
 C_{M,2N}^{\rm sing.}(1) = 
 S_0 - \frac{1}{2}
\quad .
\end{equation}
The sum $S_0$ is given in (\ref{S0}) in terms of the ratio
$\rho$ defined through
\begin{equation}
 N = \rho (M+1)
\quad .
\end{equation}
This  gives for the critical specific heat,
\begin{equation}
 C_{M,2N}^{\rm sing.}(1) =
 \frac{\ln{M}}{\pi}
 \left(
  1 + \frac{1}{M}
 \right)
 +
 c_0
 +
 \frac{c_1}{M}
 +
 \frac{c_2}{M^2}
 +
 \frac{c_3}{M^3}
 +
 {\cal{O}}
 \left(
 \frac{1}{M^4}
 \right)
\quad ,
\label{fsd}
\end{equation}
where 
\begin{eqnarray}
c_0
& = &
\frac{1}{\pi}
\left(
  \gamma_E
  +
  \frac{3\ln{2}}{2}
  -
  \ln{\pi}
  -
  2W_1(\rho)
\right)
-\frac{1}{2}
\quad ,
\\
c_1
& = &
\frac{1}{\pi}
\left(
  \gamma_E
  +
  \frac{3\ln{2}}{2}
  -
  \ln{\pi}
  +1
  -
  \frac{\pi}{4 \sqrt{2}}
  -
  2W_1(\rho)
\right)
\quad ,
\\
c_2
& = &
\frac{1}{2\pi} + \frac{\pi}{144}
+\frac{\pi}{6}W_2(\rho)
-
\frac{\rho \pi^2}{3} W_3(\rho)
\quad ,
\\
c_3
& = &
-\frac{1}{6\pi} - \frac{\pi}{144}
-\frac{\pi}{6}W_2(\rho)
+
\frac{\rho \pi^2}{3} W_3(\rho)
\quad ,
\end{eqnarray} 
where 
$\gamma_E \approx 0.577\,215\,664\,9\dots$ is the Euler-Mascheroni 
constant and where 
the functions $W_k(\rho)$ are given in the appendix.
Typical values of the constants $c_0$--$c_3$ are given in 
Table~\ref{taba}.

\noindent
\begin{table}[ht]
\caption{Values of the coefficients $c_0$--$c_3$ for 
various values of the ratio $\rho = N/(M+1)$.}
\label{taba}
\begin{center}
\begin{small}
\vspace{0.5cm}  
\noindent\begin{tabular}{|c|r|r|r|} 
    \hline 
    \hline
     $\rho$ &
\multicolumn{1}{c|}{$1/2$} &
\multicolumn{1}{c|}{$1$}  &
\multicolumn{1}{c|}{$2$}   \\[0.2cm]
    \hline
$c_0$ &  $-0.376\,674\,231\,4\dots$ &  $-0.350\,879\,731\,2\dots$ &  $-0.349\,694\,204\,8\dots$ \\
\hline 
$c_1$ &   $~~0.264\,858\,959\,5\dots$ &   $~~0.290\,653\,459\,7\dots$ &   $~~0.291\,838\,986\,0\dots$ \\
\hline 
$c_2$ &   $~~0.125\,896\,138\,1\dots$ &   $~~0.175\,784\,346\,0\dots$ &   $~~0.180\,950\,438\,8\dots$ \\
\hline 
$c_3$ &  $-0.019\,792\,842\,7\dots$ &  $-0.069\,681\,050\,6\dots$ &  $-0.074\,847\,143\,4\dots$ \\
    \hline
    \hline
  \end{tabular}
\end{small}
\end{center}
\vspace*{0.3cm}
\end{table}

So for the critical specific heat, 
apart from the $\ln{M}/M$ term, the FSS
is qualitatively 
the same as (but quantitatively different to)
that of the torus topology
(see (\ref{Salas2.13})and \cite{FF,IzHu00,Sa00}). We note that
with the Brascamp-Kunz boundary conditions, it is far easier
to extract the FSS behaviour. Indeed, determination of 
${\cal{O}}(1/{M^4})$ and higher terms
proceeds is a similarly straightforward manner.

\paragraph{Specific Heat near the Critical Point:}

The pseudocritical point, $z_{M,2N}^{\rm{pseudo}}$, 
is the value of the temperature
at which the specific heat has its maximum for a finite $M\times 2N$ lattice.
One can determine this quantity as the point where the 
 derivative of $C_{M,2N} (z)$ vanishes. 

The specific heat  is given in (\ref{ce}).
This may now be expanded about the critical point $z=1$.
This Taylor expansion simplifies using (\ref{simpl1}) and 
(\ref{simpl2}). Indeed,
\begin{eqnarray}
C_{M,2N} (z)
=
S_0-\frac{1}{2}
+
(z-1)\left[ -3S_0+1\right]
+
(z-1)^2
\left[
\frac{3}{2}S_2 + 6S_0- \frac{3}{2}
\right]
\nonumber \\
+(z-1)^3 
\left[
-5 S_2-10 S_0  +2
\right]
+{\cal{O}}\left( (z-1)^4  \right)
\quad .
\label{cexp}
\end{eqnarray}
The sums $S_0$ and $S_2$ are given in (\ref{S0}) and (\ref{S2})
respectively.

 From (\ref{cexp}), the first derivative of the
specific heat on a 
finite lattice near the infinite volume critical point is
\begin{equation}
\frac{dC_{M,2N}^{\rm sing.} (z)}{dz}
=
1-3 S_0
+3(z-1) [S_2+4S_0-1]
+3(z-1)^2 [-5S_2-10S_0+2]
+ O \left( (z-1)^3\right)
\quad .
\end{equation}
Noting that, while $S_2$ is ${\cal{O}}\left(M^2\right)$,
$S_0$ is ${\cal{O}}\left(\ln{M}\right)$, one sees that
this derivative vanishes when
\begin{equation}
 z-1
=
 \frac{3S_0-1}{3\left[S_2+4S_0-1 \right]}
+
\frac{\left[ 5S_2+10S_0-2 \right]\left[ 3S_0-1\right]^2}{
9 \left[ S_2+4S_0-1 \right]^3}
+
{\cal{O}}\left(\left(z-1 \right)^3\right)
\quad .
\label{near}
\end{equation}
Expansion of (\ref{near}) now gives the FSS of the pseudocritical point
to be
\begin{equation}
 z_{M,2N}^{\rm{pseudo}}
 =
 1
 +
 a_2 \frac{\ln{M}}{M^2}
 + 
 \frac{b_2}{M^2}
 + 
 a_3 \frac{\ln{M}}{M^3}
 + 
 \frac{b_3}{M^3}
 + 
 {\cal{O}}\left( \frac{(\ln{M})^2}{M^4} \right)
 + 
 {\cal{O}}\left( \frac{\ln{M}}{M^4} \right)
 + 
 {\cal{O}}\left( \frac{1}{M^4} \right)
\quad ,
\label{pseudocritical}
\end{equation}
where 
\begin{eqnarray}
 a_2
 & = & 
 - \frac{\pi^2}{
                2(\zeta(3)-2W_4(\rho) -4\pi \rho W_5(\rho))
               }
\quad ,
 \\
 b_2
 & = & 
 - \frac{\pi^2}{12}
   \frac{ 6 \gamma_E + 9 \ln{2}-6\ln{\pi}-2\pi -12W_1(\rho)
         }{
    \zeta(3)-2W_4(\rho)-4\pi \rho W_5(\rho)}
\quad ,
 \\
 a_3
 & = & 
 \frac{\pi^2}{\zeta(3)-2W_4(\rho)-4\pi \rho W_5(\rho)}
\quad ,
 \\
 b_3
 & = & 
 - \frac{\pi^2}{2}
 \frac{1 - 2\gamma_E -3 \ln{2}+2\ln{\pi}
 + \pi  - \pi/(4\sqrt{2})
 + 4 W_1(\rho)}{\zeta(3)-2W_4(\rho)-4\pi \rho W_5(\rho)}
\quad .
\end{eqnarray}
Again, the functions $W_k(\rho)$ are given in the appendix.
Typical values of the coefficients are given in Table~\ref{tabb}.
Thus there is no leading $1/M$ or $\ln{M}/M$ term
in the FSS of the specific heat pseudocritical point
and the specific heat shift exponent does not coincide with 
$1/\nu = 1$. This is consistent with the result (\ref{psze})
for the finite-size scaling of the real part of the Fisher zeroes.

\noindent
\begin{table}[ht]
\caption{Values of the coefficients $a_2$--$d_3^\prime$ for 
various values of the ratio $\rho = N/(M+1)$.}
\label{tabb}
\begin{center}
\begin{small}
\vspace{0.5cm}  
\noindent\begin{tabular}{|c|r|r|r|} 
    \hline 
    \hline
     $\rho$ &
\multicolumn{1}{c|}{$1/2$} &
\multicolumn{1}{c|}{$1$}  &
\multicolumn{1}{c|}{$2$}   \\[0.2cm]
    \hline                             
$a_2$ &        $-5.696\,643\,550\,4\dots$ &  $ -4.200\,054\,895\,9\dots$ &  $ -4.105\,621\,572\,1\dots$  \\
\hline 
$a_3$ &        $11.393\,287\,100\,7\dots$ &  $~~8.400\,109\,791\,7\dots$ &  $  8.211\,243\,144\,1\dots$  \\
\hline 
$b_2$ &        $ 3.758\,407\,422\,6\dots$ &  $~~2.430\,665\,892\,0\dots$ &  $  2.360\,724\,066\,8\dots$  \\
\hline 
$b_3$  &      $-16.015\,279\,517\,2\dots$ &  $-11.127\,129\,852\,1\dots$ &  $-10.846\,367\,057\,9\dots$  \\
\hline 
$c_0^\prime$  &$-0.376\,674\,233\,5\dots$ &  $ -0.350\,879\,733\,3\dots$ &  $ -0.349\,694\,207\,0\dots$ \\
\hline 
$c_1^\prime$  & $0.264\,858\,957\,4\dots$ &  $  0.290\,653\,45\,76\dots$ &  $  0.291\,838\,983\,9\dots$ \\
\hline 
$c_2^\prime$ & $ 1.309\,837\,121\,4\dots$ &  $  0.847\,424\,969\,7\dots$ &  $  0.829\,066\,715\,2\dots$ \\
\hline 
$c_3^\prime$ & $-6.557\,959\,374\,1\dots$ &  $ -4.204\,047\,250\,7\dots$ &  $ -4.086\,049\,903\,8\dots$ \\
\hline 
$d_2$ &        $-3.589\,014\,676\,5\dots$ &  $ -2.321\,114\,940\,5\dots$ &  $ -2.254\,325\,418\,5\dots$ \\
\hline 
$d_3$ &        $11.704\,450\,471\,3\dots$ &  $  8.304\,511\,256\,4\dots$ &  $  8.103\,192\,065\,9\dots$ \\
\hline 
$d_2^\prime$ &  $2.719\,946\,882\,3\dots$ &  $  2.005\,378\,462\,3\dots$ &  $  1.960\,289\,872\,9\dots$ \\
\hline 
$d_3^\prime$ & $-2.719\,946\,882\,3\dots$ &  $ -2.005\,378\,462\,3\dots$ &  $ -1.960\,289\,872\,9\dots$ \\
    \hline
    \hline
  \end{tabular}
\end{small}
\end{center}
\end{table}

Inserting (\ref{pseudocritical}) for the pseudocritical point
into (\ref{cexp})  gives the FSS behaviour of
the peak of the specific heat. This is
\begin{eqnarray}
 C_{M,2N}^{\rm sing.}(z_{M,2N}^{\rm{pseudo}}) =
& & 
 \frac{\ln{M}}{\pi}\left(1+\frac{1}{M} \right)
 +
 c^\prime_0
 +
 \frac{c^\prime_1}{M}
 +
 d_2^\prime
        \frac{ (\ln{M})^2 }{M^2}
 +
 d_2
 \frac{\ln{M}}{M^2}
 +
 \frac{c^\prime_2}{M^2}
\nonumber
\\
& & 
 +
 d_3^\prime
 \frac{(\ln{M})^2}{M^3}
 +
 d_3
 \frac{\ln{M}}{M^3}
 +
 \frac{c^\prime_3}{M^3}
+
{\cal{O}} \left(\frac{(\ln{M})^2}{M^4}\right)
\quad ,
\label{cpeak}
\end{eqnarray}
where 
\begin{eqnarray}
c^\prime_0
& = &
c_0
\quad ,
\\
c^\prime_1
& = &
c_1
\quad ,
\\
d_2^\prime
& = &
\frac{3\pi}{4}\frac{1}{\zeta(3)-2W_4(\rho)-4\pi \rho W_5(\rho)}
\quad ,
\\
d_2
& = &
\frac{\pi}{4}
\frac{ 6\gamma_E+9\ln{2}-6\ln{\pi}-2\pi-12W_1(\rho)
     }{
      \zeta(3)-2W_4(\rho)-4\pi \rho W_5(\rho)
     }
\quad ,
\\
c^\prime_2
& = &
c_2
+
\frac{\pi}{48}
\frac{
\left(
 6\gamma_E+9\ln{2}-6\ln{\pi}-2\pi-12W_1(\rho)
\right)^2
}{\zeta(3)-2W_4(\rho)-4\pi \rho W_5(\rho)} \quad ,
\\
d^\prime_3
& = &
-d^\prime_2
\quad ,
\\
d_3
& = &
\frac{\pi^2}{2 ( \zeta(3)-2W_4(\rho)-4\pi\rho W_5(\rho))}
\left(
 2 - \frac{3}{4\sqrt{2}}
\right.
\nonumber
\\
& & 
\quad \quad \quad 
\left. + \frac{3}{\pi}
\left(
 1-\gamma_E - \frac{3\ln{2}}{2}
+ \ln{\pi} + 2 W_1(\rho)
\right)
\right)
\quad ,
\\
c^\prime_3
& = &
c_3
+
\frac{\pi^3}{4 (\zeta(3)-2W_4(\rho)-4\pi\rho W_5(\rho))}
\left\{
-\frac{3}{\pi^2}
\left(
 \gamma_E
+\frac{3 \ln{2}}{2}
- \ln{\pi}
-2W_1(\rho)
\right)^2
\right.
\nonumber
\\
& & 
\left.
+ \frac{1}{\pi}
\left(
 \gamma_E+\frac{3 \ln{2}}{2}
- \ln{\pi}
-2W_1(\rho)
\right)
\left( 4 + \frac{6}{\pi}-\frac{3}{2\sqrt{2}}
\right)
-1 -\frac{2}{\pi}+\frac{1}{2\sqrt{2}}
\right\}
\quad .
\end{eqnarray} 
Again, typical values for the coefficients are compiled in Table~\ref{tabb}.

One remarks that, up to ${\cal{O}}(1/M)$, this is quantitatively
the same at the critical specific heat scaling of (\ref{fsd}).
One further remarks that the higher order structure 
of (\ref{cpeak})
differs to that at the critical point as given by 
(\ref{fsd}) in that there are  logarithmic modifications to
the subdominant terms.

\section{Conclusions}
\setcounter{equation}{0}
We have derived exact expressions for the finite-size scaling of
the Fisher zeroes for the Ising model with Brascamp-Kunz boundary 
conditions. The shift exponent, characterizing the scaling of
the corresponding pseudocritical point is $\lambda = 2$ and is
not the same as the correlation length critical exponent $\nu =1$.
This is consistent with  numerical results
for lattices with 
spherical topology \cite{HoLa96}.
Corrections to FSS can also be exactly determined and are found
to be analytic. This is consistent with the large lattice 
numerical calculations of \cite{ADH} for toroidal lattices.

A similar 
analysis applied to specific heat at criticality yields results 
qualitatively similar to those on a torus 
\cite{FF,IzHu00,Sa00,solvedIsing}.
These results complement the finite size scaling of the zeroes.
I.e., apart from the leading logarithm, only analytic corrections appear.

The finite size scaling of the pseudocritical point,
determined from the specific heat maximum is
 governed by a shift
exponent $\lambda = 2$  with logarithmic corrections. This
is again compatible with previous numerical results
\cite{GoMa93,DGS,HoLa96}. 
Finally, the first few terms for the FSS of the specific heat 
peak are
derived. 
Again all corrections are analytic, but 
in contrast to the specific heat at the critical point,
here they include logarithms.

\paragraph{Acknowledgements:}

R.K. would like to thank Wolfgang Grill for his hospitality during 
an extended
stay at Leipzig University in the framework of the {\em International 
Physics
Studies Program\/},  where this work was initiated.

\appendix
\section{Appendix}
\label{app}
\setcounter{equation}{0}

Consider, firstly, the sum
\begin{equation}
\frac{1}{N}
\sum_{i=1}^N{
              \frac{1}{
                      \left( Z-\cos{\theta_i} \right)^k
                      }
             }
\quad,
\label{1stsum}
\end{equation}
where $\theta_i = (2i-1)\pi/2N$, $k$ is a positive integer 
and $Z>1$ is independent of $n$.
We firstly treat the case $k=1$. 
We follow a similar calculation in \cite{CaPe98} and construct
\begin{equation}
 {\cal{F}}(z)
 =
 \frac{
        \cot{\pi z}
      }{
        Z-\cos{
                     \frac{
                            (2z-1)\pi
                          }{
                            2N
                          }
              }
      }
\quad .
\end{equation}
Integrate ${\cal{F}}(z)$ along the rectangular contour bounded by 
${\rm{Re}}z = 1/2+\epsilon$,
${\rm{Re}}z = 2N+ 1/2+\epsilon$
and ${\rm{Im}}z = \pm i R$ where $R$ is some large real constant and 
$\epsilon$ is a small real number which we take to be positive.
Because of the periodic nature of the integrand, the integrals along
the left and right edges of the contour cancel. Further, due to
the $\cot{\pi z}$ term, the integrand
and hence the full integral vanishes in the limit of infinitely large
$R$. 
Now, ${\cal{F}}(z)$ has $2N$ simple
poles inside the contour along the real 
axis and a further two coming from the simple zeroes of its
denominator.
The residue theorem then gives the exact result
\begin{equation}
\frac{1}{N}
\sum_{i=1}^N{
              \frac{1}{
                       Z-\cos{\theta_i}
                      }
             }
=
\frac{
1
}{\sqrt{Z^2-1}}
       \tanh{
              \left(
                     N \cosh^{-1}{Z}
             \right)
          }
\quad .
\label{Ia}
\end{equation}
This result is also implicitly contained in \cite{JaKl90}.
Differentiation of (\ref{Ia}) with respect to $Z$ gives the sum
(\ref{1stsum}) with larger values of $k$. In particular, we also
find
\begin{equation}
\frac{1}{N}
\sum_{i=1}^N{
              \frac{1}{\left(
                       Z-\cos{\theta_i}
                      \right)^2}
             }
=
\frac{Z \tanh{\left(
 N \cosh^{-1}{Z}
\right)}
}{(Z^2-1)^{3/2}}
-
\frac{N {\rm{sech}}^2{\left(
 N \cosh^{-1}{Z}
\right)}
}{Z^2-1}
\quad .
\label{Ib}
\end{equation}

The purpose of the remainder of this appendix is to calculate the
two sums $S_0$ and $S_2$ where 
\begin{equation}
 S_k = \frac{1}{M} \sum_{j=1}^M g_j^{(k)}(1)
\quad ,
\label{Sk}
\end{equation}
where
$g_j^{(k)}(z) = d^k g_j(z)/dz^k$,
\begin{equation}
 g_j(z)
 =
 \frac{
       \tanh{
             \left(
                    N
                    \cosh^{-1}{
                               (1/z+z-\cos{\phi_j})
                              }
             \right)
            }
     }{
       \sqrt{
             (1/z+z-\cos{\phi_j})^2-1
            }
     }
\quad ,
\end{equation}         
 and where $\phi_j = \pi j / (M+1)$.
It is convenient to rewrite $g_j(z)$ as
\begin{eqnarray}
\lefteqn{
g_j(z)
= 
{
               \frac{
                       1
                    }{
                       \sqrt{
                             (2+m^2-\cos{\phi_j})^2-1
                            }
                    }
             }
}\nonumber \\
& & 
- 2
 {
               \frac{
                     1
                     }{
                       \sqrt{(2+m^2-\cos{\phi_j})^2-1}
                     }
             }
\left\{
        \exp{
             \left[
                   2N\cosh^{-1}{
                                 (2+m^2-\cos{\phi_j})
                                }
            \right]
            }
        +1
\right\}^{-1}
\quad ,
\label{bigsum}
\end{eqnarray}
where $m^2=1/z+z-2$.
We consider, firstly,  the sum
\begin{equation}
\sum_{j=1}^M{
             \frac{1}{
                      \sqrt{
                            \left(
                                  2+m^2
-\cos{
                                              \phi_j
                                        }
                            \right)^2
                            -1
                           }
                      }
             }
\quad,
\label{CP}
\end{equation}
and we are interested in the limit $m \rightarrow 0$.
This may be calculated by direct application of the Euler-Maclaurin 
formula \cite{Abramowitz}. 
In fact, the more general summation (\ref{CP}) has been
calculated in \cite{CaPe98}. Writing $m = \eta/(M+1)$, that
sum is given by
\begin{eqnarray}
\lefteqn{
\frac{1}{M+1}
\sum_{j=1}^M{
             \frac{1}{
                      \sqrt{
                            \left(
                                  2+\left(\frac{\eta}{M+1} \right)^2
-\cos{
                                              \phi_j
                                        }
                            \right)^2
                            -1
                           }
                      }
             }
=
}
& & 
\nonumber
\\
& &
\frac{\ln{(M+1)}}{\pi}
+
\frac{1}{\pi}
\left[
 \gamma_E - \ln{\pi} + \frac{3 \ln{2}}{2}
 + 
 G_0\left( \frac{\sqrt{2} \eta}{\pi} \right)
\right]
-
\frac{1}{M+1} \frac{1}{4 \sqrt{2}}
-
\frac{\ln{(M+1)}}{(M+1)^2} \frac{\eta^2}{4 \pi}
\nonumber \\
& & 
+ 
\frac{1}{2}
\frac{1}{(M+1)^2}
\left[
 \frac{\pi}{6}
 \left(
            \frac{1}{12} - G_1\left( \frac{\sqrt{2} \eta}{\pi} \right)
 \right)
+ \frac{\eta^4}{3 \pi^3} H_1\left( \frac{\sqrt{2} \eta}{\pi} \right)
-\frac{\eta^2}{2 \pi}
\left(
 \gamma_E - \ln{\pi} + \frac{3 \ln{2}}{2}
\right.
\right.
\nonumber \\
& &
\left.
\left.
 + 
 \frac{2}{3}G_0\left( \frac{\sqrt{2} \eta}{\pi} \right)
 - \frac{1}{3}
\right)
\right]
+ 
\frac{1}{(M+1)^3}
\frac{3 \sqrt{2}}{64}
\eta^2
+ 
{\cal{O}}\left(
\frac{\ln{M}}{(M+1)^4}
\right)
+ 
{\cal{O}}\left(
\frac{1}{(M+1)^4}
\right)
\quad,
\label{IIa}
\end{eqnarray}
where $\gamma_E \approx 0.577\,215\,664\,9\dots$ is the Euler-Mascheroni 
constant \cite{Abramowitz} and 
where the remnant functions are
\begin{eqnarray}
 G_0 (\alpha)
 & = & 
 \sum_{n=1}^\infty{
 (-1)^n \left(
 \begin{array}{c} 2n \\ n \end{array}
 \right)
 \zeta(2n+1)
 \left( \frac{\alpha}{2}\right)^{2n}
 }
\quad ,
\\
 G_1 (\alpha)
 & = & 
 2
 \sum_{n=1}^\infty{
 \frac{(-1)^n}{n} \left(
 \begin{array}{c} 2n \\ n \end{array}
 \right)
 \zeta(2n+1)
 \left( \frac{\alpha}{2}\right)^{2n+2}
 }
\quad ,
\\
 H_1 (\alpha)
 & = & 
 \sum_{n=1}^\infty{
 (-1)^n (2n+1)
 \left(
 \begin{array}{c} 2n \\ n \end{array}
 \right)
 \zeta(2n+3)
 \left( \frac{\alpha}{2}\right)^{2n}
 }
\quad ,
\end{eqnarray}
in which $\zeta(n)$ is a Riemann zeta function \cite{Abramowitz},
\begin{equation}
 \zeta(n) =
 \sum_{k=1}^\infty{\frac{1}{k^n}}
\quad .
\end{equation}
Setting $\eta = 0$, one has the first component of the sum 
appearing in $S_0$,
\begin{eqnarray}
\lefteqn{
\frac{1}{M+1}
\sum_{j=1}^M{
             \frac{1}{
                      \sqrt{
                            \left(
                                  2
-\cos{
                                              \phi_j
                                        }
                            \right)^2
                            -1
                           }
                      }
             }
=
\frac{\ln{(M+1)}}{\pi}
}
& & 
\nonumber
\\
& &
+
\frac{1}{\pi}
\left[
 \gamma_E - \ln{\pi} + \frac{3 \ln{2}}{2}
\right]
-
\frac{1}{M+1} \frac{1}{4 \sqrt{2}}
+ 
\frac{1}{(M+1)^2}
 \frac{\pi}{144}
+ 
{\cal{O}}\left(
\frac{1}{(M+1)^4} 
\right)
\quad.
\label{IIb}
\end{eqnarray}
Next, we consider
\begin{equation}
\sum_{j=1}^M{
               \frac{
                     1
                     }{
                       \sqrt{\left(2+\left( \frac{\eta}{M+1} \right)^2
-\cos{\phi_j}\right)^2-1}
                     }
             }
\left\{
        \exp{
             \left[
                   2N\cosh^{-1}{
                                 (2-\cos{\phi_j})
                                }
            \right]
            }
        +1
\right\}^{-1}
\equiv
\sum_{j=1}^M{
 h_j(\eta)
}
\quad .
\label{dem}
\end{equation}
It is natural to define an aspect ratio, $\rho$, through
\begin{equation}
 N = \rho (M+1)
\quad ,
\end{equation}
and introduce
\begin{eqnarray}
X_j & = & \sqrt{2 \eta^2+\pi^2j^2}
\quad ,
\\
Y_j & = & \exp{(2 \rho X_j)} + 1
\quad .
\end{eqnarray}
Dominant contributions to (\ref{dem}) 
come from the small $j$ terms. 
The sum may thus
replaced by \cite{CaPe98}
\begin{equation}
\sum_{j=1}^\infty{
 h_j(\eta)
}
\quad ,
\label{rep}
\end{equation}
where the expansion of the summand is
\begin{equation}
 h_j(\eta)
=
 \frac{M+1}{X_j Y_j}
 +
 \frac{1}{M+1}
 \left\{
 \frac{1}{24} \frac{\pi^4 j^4}{x_j^3 Y_j}
 -
 \frac{X_j}{8 Y_j}
 + 
 \frac{\rho}{12}
 \frac{Y_j-1}{Y_j^2}
 \left(
 \frac{\pi^4 j^4}{X_j^2}+X_j^2
 \right)
 \right\}
+ {\cal{O}}\left(\frac{1}{(M+1)^3} \right)
\quad .
\label{IIc}
\end{equation}
Taking the limit of this quantity as $\eta \rightarrow 0$ gives
the second
component of the sum 
appearing in $S_0$,
\begin{eqnarray}
\lefteqn{
\frac{1}{M+1}\sum_{j=1}^M{
               \frac{
                     1
                     }{
                       \sqrt{(2
-\cos{\phi_j})^2-1}
                     }
             }
\left\{
        \exp{
             \left[
                   2N\cosh^{-1}{
                                 (2-\cos{\phi_j})
                                }
            \right]
            }
        +1
\right\}^{-1}
=
}
& & 
\nonumber \\
& & 
\frac{1}{\pi} W_1(\rho) 
+
\frac{1}{(M+1)^2}
\left[
 -\frac{\pi}{12}W_2(\rho) 
 + 
\frac{\rho \pi^2}{6}W_3(\rho) 
\right]
+
{\cal{O}}\left(
 \frac{1}{(M+1)^4}
\right)
\quad ,
\label{IId}
\end{eqnarray}
where
\begin{eqnarray}
W_1 (\rho) & = & \sum_{n=1}^\infty{\frac{1}{n (e^{2 \rho \pi n}+1)}}
\quad ,\\
W_2 (\rho) & = & \sum_{n=1}^\infty{\frac{n}{ (e^{2 \rho \pi n}+1)}}
\quad ,\\
W_3 (\rho) & = & \sum_{n=1}^\infty{\frac{n^2e^{2 \rho \pi n}}{
(e^{2 \rho \pi n}+1)^2}}
\quad .
\end{eqnarray}
The functions $W_k(\rho)$ are rapidly converging sums which may be computed
numerically.  For example,
at $\rho = 1$,
$W_1 \approx 0.001\,865\,707\,7\dots$,
$W_2 \approx 0.001\,870\,956\,1\dots$
and 
$W_3 \approx 0.001\,874\,495\,5\dots$.
Finally, from (\ref{Sk}), (\ref{bigsum}), (\ref{IIb})
and  (\ref{IId}), the desired result is
\begin{eqnarray}
\lefteqn{
          S_0 = \frac{\ln{M}}{\pi}
          + 
          \frac{1}{\pi}
          \left(
                \gamma_E
                  +
                \frac{3\ln{2}}{2}
                -
                \ln{\pi}
                -
                2W_1(\rho)
           \right)
           +
           \frac{\ln{M}}{M}
           \frac{1}{\pi}
}
& &
\nonumber
\\
& & 
+ \frac{1}{M}
\frac{1}{\pi}
\left(
  \gamma_E
  +
  \frac{3\ln{2}}{2}
  -
  \ln{\pi}
  +1
  -
  \frac{\pi}{4 \sqrt{2}}
  -
  2W_1(\rho)
\right)
\nonumber
\\
& & +
\frac{1}{M^2}
\left(
\frac{1}{2\pi} + \frac{\pi}{144}
+\frac{\pi}{6}W_2(\rho)
-
\frac{\rho \pi^2}{3} W_3(\rho)
\right)
\nonumber
\\
& & 
-\frac{1}{M^3}
\left(
\frac{1}{6\pi} + \frac{\pi}{144}
+\frac{\pi}{6}W_2(\rho)
-
\frac{\rho \pi^2}{3} W_3(\rho)
\right)
+ {\cal{O}}\left(
       \frac{1}{M^4}
\right)
\quad .
\label{S0}
\end{eqnarray} 

The sum  $S_2$ is 
\begin{eqnarray}
S_2 &=& \frac{d^2}{dz^2}
      \frac{1}{M}
      \left. \sum_{j=1}^M{g_j(z)} \right|_{z=1}
\\
& =&
  \frac{d^2}{dz^2}
      \frac{1}{M}
      \left.  \sum_{j=1}^M{
\frac{
    1 - 2 \left\{
                \exp{\left[
                     2 N \cosh^{-1}{(1/z+z-\cos{\phi_j})}
                    \right]}+1
          \right\}^{-1}
}{\sqrt{(1/z+z-\cos{\phi_j})^2-1}}
} \right|_{z=1} \quad .
\end{eqnarray}
The sums involved here are found by taking appropriate derivatives
of  (\ref{IIa}) and (\ref{IIc}).
The result for $S_2$ is

\begin{eqnarray}
\lefteqn{ S_2
 =
 M^2
 \frac{-2}{\pi^3}
 \left[
       \zeta(3) - 2 W_4(\rho) - 4 \pi \rho W_5(\rho)
 \right]
\nonumber
}
& &
\nonumber 
\\
& & 
+
 M \frac{-6}{\pi^3}
 \left[
       \zeta(3) - 2 W_4(\rho) - 4 \pi \rho W_5(\rho)
 \right]
-
\frac{\ln{M}}{2 \pi}
+
{\cal{O}} \left(1 \right)
\quad ,
\label{S2}
\end{eqnarray}
in which
$\zeta(3) \approx 1.202\,056\,903\,2\dots$ and
\begin{eqnarray}
 W_4(\rho) & = & \sum_{n=1}^\infty{\frac{1}{n^3(\exp{(2\rho \pi n)}+1)}}
\quad ,
\\
 W_5(\rho) & = & \sum_{n=1}^\infty{\frac{\exp{(2\rho \pi n)}}{
n^2(\exp{(2\rho \pi n)}+1)^2}}
\quad .
\end{eqnarray}
Typical values of these rapidly converging sums are 
$W_4 \approx 0.001\,864\,398\,1\dots$ and
$W_5 \approx 0.001\,861\,360\,1\dots$
at $\rho = 1$.



\end{document}